\documentclass[useAMS,usenatbib]{mn2e}
\usepackage{times}
\usepackage{rotating}
\usepackage{amsmath}
\usepackage{amssymb}
\usepackage[dvipsnames]{xcolor}
\usepackage{graphicx}
\usepackage{graphics}
\usepackage{hyperref}
\hypersetup{
     colorlinks   = true,
     citecolor    = blue
}

\long\def\symbolfootnote[#1]#2{\begingroup%
\def\thefootnote{\fnsymbol{footnote}}\footnote[#1]{#2}\endgroup}
\newcommand{\gae}{\lower 2pt \hbox{$\, \buildrel {\scriptstyle >}\over {\scriptstyle \sim}\,$}}
\newcommand{\lae}{\lower 2pt \hbox{$\, \buildrel {\scriptstyle <}\over {\scriptstyle \sim}\,$}}
\newcommand{\aprop}{\lower 2pt \hbox{$\, \buildrel {\scriptstyle \propto}\over {\scriptstyle \sim}\,$}}

\def\d{\mathrm{d}}

\def\nGJ{\mbox{\large n}_{_{GJ}}}

\def\n13{\mbox{\large n}_{_{13}}}
\def\nk{\mbox{{\large n}}}

\def\bk{{\bf k}}
\def\alfven{Alfv\'en }

\begin{document}

\title[FRB radiation and \alfven waves]
{FRB Coherent Emission from Decay of \alfven Waves}

\author[Kumar]{Pawan Kumar\thanks{E-mail:
    pk@astro.as.utexas.edu,
  zmbosnjak@gmail.com}$^1$ and \v Zeljka Bo\v snjak$^{2}$
\\ $^{1}$Department of Astronomy, University of Texas at Austin, Austin,
 TX 78712, USA\\
$^{2}$Faculty of Electrical Engineering and Computing, University of Zagreb, 10000 Zagreb, Croatia}


\maketitle

\begin{abstract}
We present a model for FRBs where a large amplitude 
\alfven wave packet is launched by a disturbance near the surface of a    
magnetar, and a substantial fraction of the wave energy
is converted to coherent radio waves at a distance of a few tens of neutron
star radii. The wave amplitude at the magnetar surface should be about 
10$^{11}$G in order to produce a FRB of isotropic luminosity 10$^{44}$ 
erg s$^{-1}$. An electric current along the static magnetic field is required 
by \alfven waves with non-zero component of transverse wave-vector. 
The current is supplied by counter-streaming electron-positron pairs, 
which have to move at nearly the speed of light at larger radii as the 
plasma density decreases with distance from the magnetar surface. 
The counter-streaming pairs are subject to two-stream instability 
which leads to formation of particle bunches of size of order 
$c/\omega_p$; where $\omega_p$ is plasma frequency. 
A strong electric field develops along the static magnetic field when the 
wave packet arrives at a radius where electron-positron density is 
insufficient to supply the current required by the wave.
The electric field accelerates particle bunches along the curved magnetic
field lines, and that produces the coherent FRB radiation. We provide a 
number of predictions of this model. 

\end{abstract}

\begin{keywords}
Radiation mechanisms: non-thermal - methods: analytical - stars: magnetars
- radio continuum: transients - masers
\end{keywords}

\section{Introduction}

It is firmly established that at least a few milli-second duration, very bright,
 radio signals that have been detected between about 400 MHz and 7 GHz 
(known as Fast Radio Bursts or FRBs) are coming from a distance of about 
a Gpc or more, eg. Lorimer et al. 2007; Thornton et al. 2013; 
Spitler et al. 2014; Petroff et al. 2016; Bannister et al. 2017; 
Law et al. 2017; Chatterjee et al. 2017; Marcote et al. 2017; 
Tendulkar et al. 2017; Gajjar et al. 2018; Michilli et al. 2018; 
Farah et al. 2018; Shannon et al. 2018; Oslowski et al. 2019; 
Kocz et al. 2019; Bannister et al. 2019; CHIME Collaboration 2019a and 2019b; 
Ravi 2019a, 2019b; Ravi et al. 2019.

Many mechanisms have been suggested for the generation of high luminosity
coherent radio waves from Fast Radio Bursts (FRBs), eg. Katz (2014, 2016),
Murase et al. (2016; 2017), Kumar et al. (2017), Metzger et al. (2017), 
Zhang (2017), Beloborodov (2017), 
Cordes (2017), Ghisellini \& Locatelli (2018), Lu \& Kumar (2018), 
Metzger et al. (2019), Thompson (2019), Wang \& Lai (2019), Wang et al. (2019);
 for a recent review see Katz (2018). 
Many of these models are severely constrained because of large
optical depth for induced Compton scatterings, and radiative acceleration
of particles in the vicinity of the source. These models, when they can
be made to work, require fine turning of parameters or very low efficiency 
for producing radio waves. 

\begin{figure*}
\vskip -2cm
  \includegraphics[width=19.3cm,height=14.53cm]{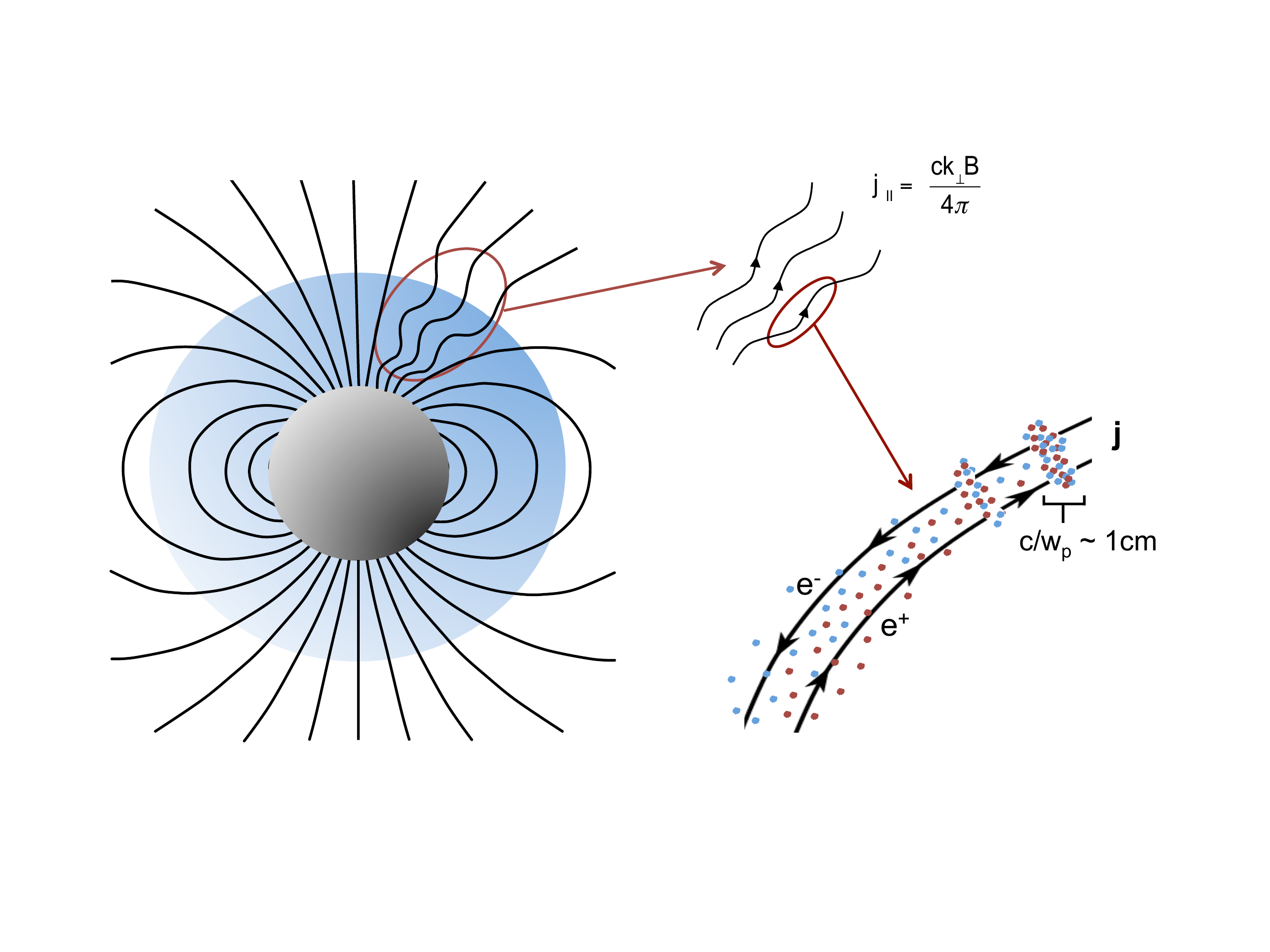}
\vskip -2.4cm
\caption{This figure provides a schematic overview of the scenario 
described in this work.
A large amplitude \alfven wave, with magnetic field perturbation of 
$B\sim 10^{11}$G, is produced by some disturbance in the magnetic pole region 
at the surface of a magnetar (the left-center sketch). As the
wave travels to larger radius, outside the shaded region, it becomes 
charge starved, i.e. $e^\pm$ density falls below a threshold value of order
10$^{13} (B_{11}/R_7^3)$ cm$^{-3}$, and a strong electric field with a 
non-zero component along the static magnetic field develops. The \alfven 
wave packet has a non-zero current, $j_\parallel$, associated with it that 
is parallel to the large scale static magnetic field (right, top panel). 
The current is carried by counter-streaming $e^\pm$ pairs, which is subject
to the 2-stream instability and that causes formation of particle clumps
of radial width $\sim c/\omega_p\sim 1$ cm (right, bottom part of the sketch).
When these clumps enter the charge starvation region, they are
accelerated by the electric field to Lorentz factor of $10^3$.
Clumps of particles moving along the curved magnetic field lines 
produce powerful coherent curvature radiation in the radio band. We find the
efficiency of converting \alfven wave energy to FRB radiation is of order
unity.
}
\label{fig-scenario}
\end{figure*}

The main components of the scenario we explore in this work are described
in Figure \ref{fig-scenario}. A high amplitude \alfven wave, 
$B/B_0\sim 10^{-4}$, is launched from the surface of a neutron 
star (NS) endowed with a strong magnetic field ($B_0\gae 10^{15}$G). 
\alfven waves have a non-zero electric current along the magnetic 
field lines unless the wave vector is exactly parallel to the field line.
The current 
is carried by electrons and positrons moving in opposite directions. 
This counter-streaming motion of e$^\pm$ is subject to 
two-stream instability, which leads to formation of roughly charge neutral
particle clumps. As the plasma density decreases with distance from the NS
surface, the particle velocity increases in order to carry the current density 
required by the \alfven wave. When the plasma density at some height in the NS 
magnetosphere falls below a critical value, the plasma is unable to support
the current even if e$^\pm$ move at the speed of light. A strong
time-dependent electric field then develops, and the displacement current
associated with the field makes up for the deficit in plasma current density.
The electric field has a component along the static magnetic field just
as the plasma current does. This electric field forces electrons and positrons 
in the clump to move at a high Lorentz factor along magnetic field lines 
in opposite directions. The coherent curvature radiation produced by these
clumps is observed as the FRB signal.

In the next section we describe the basic properties of \alfven wave
propagation through a stratified medium, the development of a
strong electric field when the wave becomes charge starved, formation
and acceleration of particle clumps. The emission of coherent radiation
and predictions of the model are discussed in \S3. The main results and 
limitations of this investigation are summarized in \S4.

\section{Alfv\'en waves dissipation and coherent radio emission}
\label{alf}

The process by which \alfven waves might be generated at the surface of 
a neutron star (NS) is not well understood. It is possible that these
waves are generated by sudden crustal motion, crustal-quake or emergence of 
magnetic flux tubes from below the surface of a NS.
We assume, as have previous work on this subject e.g. Blaes et al. 1989,
that \alfven waves are launched by some mechanism close to the surface
of a NS and study their propagation and eventual decay to radio waves.

We describe a few basic properties of \alfven waves and their propagation
in an inhomogeneous medium in the next sub-section, which are needed for 
this work. Charge starvation of \alfven waves, generation of electric fields, 
and particle acceleration are presented in \S\ref{c-starv}. Particle 
bunching due to two-stream instability associated with the \alfven wave 
current, and acceleration of bunches are discussed in \S\ref{clump-acc}.

\subsection{\alfven waves}
\label{alfven}

An \alfven wave-packet launched at the surface of a NS with amplitude
$B$ has isotropic equivalent of luminosity $L_{aw} = 
B^2 R_{ns}^2 c\sim 3{\rm x}10^{44} B_{11}^2$ erg s$^{-1}$. Therefore, 
close to the NS $B\gae 10^{11}$G is required in order for the \alfven wave 
to produce GHz EM waves at the typical FRB luminosity of $\sim 10^{44}$ 
erg s$^{-1}$ with an efficiency of order unity (see \S\ref{rad-physics}). 
If the NS has magnetar
strength magnetic field, i.e. $B_0\sim 10^{15}$G, then \alfven waves 
at the NS surface have dimensionless amplitude $B/B_0 \sim 10^{-4}$,
and they can be treated as linear perturbation in spite of their 
enormous luminosity. 

We first consider \alfven waves in a uniform magnetic 
field using linear perturbation analysis, and subsequently generalize the
discussion to an arbitrary magnetic field geometry. The physics of \alfven
waves propagation in a uniform field is straightforward, which can be
found in any introductory plasma/MHD text book such as Kulsrud (2005). We
include a brief derivation of basic properties of \alfven waves to establish 
the notation and to emphasize one particular feature which is that these
waves in general have a non-zero current parallel to the static magnetic 
field. This is one of the key properties we exploit for their dissipation 
and that leads to generation of coherent curvature radiation.

The uniform magnetic field of magnitude $B_0$ is taken to be along 
${\bf\hat z}$. The perturbation to the field is $B {\bf\hat x}$,
and $B\propto \exp(i{\bf r\cdot k_{aw}} - i\omega_{aw} t)$; where
the wave-vector ${\bf k_{aw}}$ is in some arbitrary direction.
The linearized flux freezing, fluid momentum, and induction equations are
\begin{equation}
   {\bf E}_\perp = -{{\bf v}\over c}\times {\bf B_0}, \; {\partial\rho_e{\bf v}\over
   \partial t} = { {\bf B_0\cdot\nabla {\bf B}}\over 4\pi}, \; 
   {\bf \nabla\times(v\times B_0)} = {\partial {\bf B}\over \partial t}, 
  \label{aw-eq}
\end{equation}
where $\rho_e = \rho_0 + B_0^2/(4\pi c^2)$, and $\rho_0$ is the plasma density.
These equations can be combined to obtain relationships between ${\bf v}$, 
${\bf E}_\perp$ \& ${\bf B}$, and the \alfven wave dispersion relation
\begin{equation}
  {{\bf v_\perp}\over c} = -{ {\bf B}\over B_0}, \quad {\bf E}_\perp = {\bf 
B\times\hat B_0}, \quad  ({\bf k_{aw}\cdot \hat B_0})^2 V_A^2 = \omega_{aw}^2,
   \label{aw-rel}
\end{equation}
here ${\bf v_\perp}$ is fluid velocity vector perpendicular to the 
static magnetic field, and $V_A^2 =  c^2 B_0^2/(4\pi\rho_0 c^2 + B_0^2)$ is
very close to $c^2$ in NS magnetosphere. The first part of equation 
(\ref{aw-rel}) shows that particles move with the magnetic field in the 
transverse direction like beads on a wire. In other words, particles stay 
in the ground state of Landau level all the time and their non-zero 
transverse velocity is simply to keep up with the perturbed magnetic field 
lines of the \alfven wave to which they are locked. 

The current associated with the 
wave can be calculated using the linearized Ampere's equation
\begin{equation}
  i c\, {\bf k_{aw}\times B} = 4\pi {\bf j} - i\omega_{aw} 
      ({\bf E}_\perp + {\bf E_d})
   \label{Eda}
\end{equation}
where we have explicitly decomposed the electric field as sum of the 
${\bf v\times B_0}$ field given by equation (\ref{aw-eq}), and a field
${\bf E_d}$ which provides displacement current. Making use of 
equation (\ref{aw-rel}) to eliminate ${\bf E}_\perp$ in terms of 
${\bf B}$ simplifies the above equation 
\begin{equation}
  ic {\bf k_{aw\perp}}\times {\bf B} = 4\pi {\bf j} 
  \,  - \, i\omega_{aw} {\bf E_d},
    \label{cur-a}
\end{equation}
where ${\bf k_{aw\perp}}$ is the component of ${\bf k_{aw}}$ perpendicular
to the static field ${\bf B_0}$ {\it and} the perturbed field ${\bf B}$, i.e. 
$k_{aw\perp} = {\bf k_{aw}}\cdot({\bf\hat B_0\times\hat B})$.
The left side of the equation drives the system to have non-zero current 
along ${\bf\hat B_0}$, and possibly a non-zero ${\bf E_d}$, provided
that ${\bf k_{aw}}$ does not lie in the ${\bf B_0}$--${\bf B}$ plane. 
The condition for a non-zero $E_d$ can be obtained by
defining a critical plasma density
\begin{equation}
   n_c \equiv {k_{aw\perp} B \over 8\pi q} = (10^{16}{\rm cm}^{-3}) 
      { B_{11}\over \lambda_{aw\perp,4}},
   \label{n-cs-a}
\end{equation}
in terms of which equation (\ref{cur-a}) can be recast in the following form
\begin{equation}
   {\omega_{aw} {\bf E_d} \over 8\pi qc} = i n ({\bf v_+ - v_-})/2c 
       - n_c {\bf\hat B_0}  
    \label{cur-b}
\end{equation}
where ${\bf v_+}$ (${\bf v_-}$) is the velocity of positrons (electrons),
and $n$ is the plasma density at radius $R$. As long as
$n > n_c$, the plasma has the ability to supply the current needed 
by the \alfven wave with ${\bf k_{aw\perp}}\times{\bf B}) \not =
 0$. In this case, ${\bf E_d}=0$ and $|v_\pm|/c = n_c/n$. However, when 
$n(R) < n_c(R)$, it is the displacement
current, $\partial {\bf E_d}/\partial t$, along ${\bf\hat B_0}$
that must support \alfven wave's non-zero curl of ${\bf B}$; 
this is defined as the charge starvation region. The
electric field in the charge starvation zone, in the limit of 
vanishingly small plasma density, is
\begin{equation}
   {\bf \hat B_0\cdot E_d} = -{8\pi q c n_c\over \omega_{aw}} = -{k_{aw\perp}
      \over k_{aw\parallel}} B.
   \label{Ed-B}
\end{equation}

The critical density is zero for $k_{aw\perp} = 0$ (eq. \ref{n-cs-a}). However, 
for an \alfven wave packet launched from a region of finite size, it is
impossible that $k_{aw\perp} = 0$ everywhere.  
The transverse size of the region from which the \alfven wave is launched 
cannot be larger than NS radius ($\sim10^6$cm), and is likely of order 
the size of polar cap region ($\ell_{pc}$) of open field lines that 
extend outside the light-cylinder.  Thus, we expect 
$\lambda_{aw\perp} \sim \ell_{pc}$. It is easy to show that for a NS 
with spin velocity of $\Omega_{ns}$, the size of the polar cap is 
$\ell_{pc} \sim (R_{ns}^3\Omega_{ns}/c)^{1/2} \sim 5{\rm x} 10^3 
\Omega_{ns}^{1/2}$cm.  
If the \alfven wave frequency is $10^5$Hz, then 
$k_{aw\perp}/k_{aw\parallel} \sim 30$ at the surface of the NS. 

\begin{figure}
\centerline{\hbox{\includegraphics[width=9cm, height=13cm, angle=0]{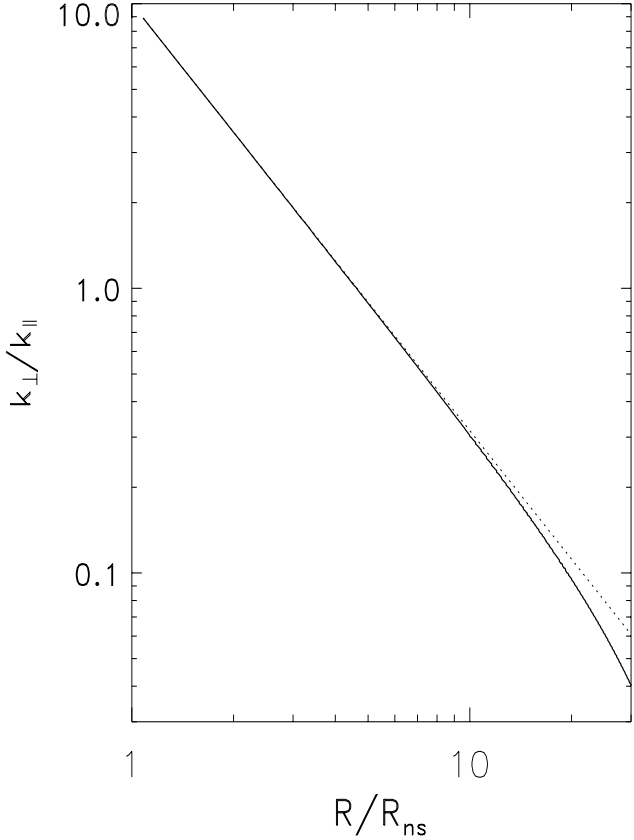}}}
\vskip -0.7cm
\caption{Shown here is the component of wave-vector \bk ~ perpendicular 
to the local magnetic field, i.e. ${\bf\hat B}_0\times{\bf k}$; the
magnetic field is taken to be dipolar. The distance
is measured in unit of neutron star radius, and the wave-number is in units
of $\omega_{aw}/c$; where $\omega_{aw}$ is \alfven wave frequency.
Shown also is $R^{-3/2}$ curve (dotted line) for comparison.
}
\label{fig1}
\end{figure}

 The \alfven wave-packet 
travels along magnetic field lines and its amplitude decreases as the field 
lines diverge or flareup with distance; $B \propto R^{-3/2}$ to conserve 
luminosity\footnote{The magnetic field decreases with distance as 
$\sim R^{-3}$, and so $B/B_0 \propto R^{3/2}$, i.e. the wave becomes
increasingly more nonlinear with distance from its launching site.}. 
Therefore, the
critical density $n_c \propto R^{-3/2} k_{aw\perp}$ (see eq. \ref{n-cs-a}). 
The component of \alfven wave-vector along $\hat{\bf B}_0$, $k_{aw\parallel}$,
is conserved as it is equal to $\omega_{aw}/c$. The wave-vector component 
perpendicular to the magnetic field $k_{aw\perp}$, however, decreases with
$R$, approximately as $R^{-3/2}$, since the transverse size of the 
wave packet increases as $R^{3/2}$ due to the spreading of magnetic
field lines. The precise behavior of $\bk$ with distance is determined 
by solving the following Eikonal equations, e.g. (Weinberg, 1962) 
\begin{equation}
  {d {\bf x}\over dt} = c {\bf\hat B_0}, \quad {d k^i_{aw}\over dt} = - 
    c\sum_{j=1}^3 k^j_{aw} {\partial {\hat B}_{0j}\over \partial x_i}.
\end{equation}
The solution of this equation is shown in Fig. \ref{fig1} for a dipole 
magnetic field, and it is clear that the component of ${\bf k_{aw}}$
transverse to the local magnetic field direction does indeed decrease 
approximately as $R^{-3/2}$. Therefore, $k_{aw\perp}/k_{aw\parallel}
\propto R^{-3/2}$ and it becomes of order unity at the radius where the
wave becomes charge starved.

The critical density as a function of $R$ is given by
\begin{equation}
     n_c(R) \equiv {k_{aw\perp} B \over 8\pi q} \approx (10^{16}{\rm cm}^{-3})
      { B_{11}\over \lambda_{aw\perp,4}}\left[ {R_{ns}\over R} \right]^3,
   \label{n-cs-b}
\end{equation}
where the factor $B_{11}/\lambda_{aw\perp,4}$ is evaluated at the NS surface.
Since $n_c\propto R^{-3}$, the current along ${\bf\hat{B}_0}$ required by 
the \alfven wave packet can be supplied by the magnetosphere plasma provided 
that particle density $n(R)$ does not decrease faster than $R^{-3}$. We 
address this next.

The particle density in the NS magnetosphere is controlled by the 
magnetic field and the rotation rate (Goldreich \& Julian, 1969). 
Charge particles with a minimum number density $\nGJ$ (Goldreich-Julian
density) fill the magnetosphere. Otherwise, the electric field is so strong
as to pull charge particles from NS surface or create $e^\pm$ spontaneously.
The actual particle density is taken to be a factor ${\cal M}$ larger than the
minimum required density:
\begin{multline}
    n = {\cal M}\, \nGJ = { {\cal M}\, {\bf B_0\cdot\Omega_{ns}}\over 2\pi qc}
    \approx {{\cal M}\, B_{ns}\Omega_{ns}\over 2\pi qc}\left( {R_{ns}\over R}
    \right)^3 \\
    \approx 10^{13}{\rm cm}^{-3} {\cal M}\, B_{ns,15} \Omega_{ns}
     \left( {R_{ns}\over R} \right)^3.
     \label{nR}
\end{multline}
The value of ${\cal M}$ is highly uncertain and is estimated to be anywhere
between a few and $\sim 10^6$ for pulsars. Therefore, it is possible that
${\cal M}$ is a function of $R$, and that $n={\cal M}\, \nGJ\propto {\cal M}
R^{-3}$ declines faster than $R^{-3}$ at some radius.

Comparing the critical density (eq. \ref{n-cs-a}) with the particle density 
in NS magnetosphere (eq. \ref{nR}), we find that $n_c > n$ unless 
${\cal M}\gae 10^3$. 
So, ${\cal M}$ should be sufficiently large near the \alfven wave launching 
site to ensure that wave is not charge starved at birth. As long as this
condition is satisfied, the wave-packet will travel to larger radius without
becoming charge starved until ${\cal M}$ starts to decrease.
If at some radius $R_c$ the particle density $n$ falls below $n_c$ due to the
decrease of ${\cal M}$, then the wave becomes charge starved and a 
strong electric field with a component parallel to the static magnetic field 
develops. The radius where the charge starvation sets in is likely not very
far away from the NS since $n_c \gg \nGJ$, so ${\cal M}$ only needs to
drop a bit for the wave to enter the charge starvation regime.
The development of electric field and particle acceleration when \alfven waves 
become charge starved is discussed in the next sub-section.

\subsection{Charge starvation and particle acceleration}
\label{c-starv}

The \alfven wave-packet encounters decreasing particle density as it 
propagates out to larger radius. When the ambient medium density 
approaches the critical density ($n_c$; eq. \ref{n-cs-a}), charge 
particles begin to move at close to the speed of light to supply the 
current needed by the \alfven wave. 
An electric field must develop when $n$ falls below $n_c$, because
the plasma cannot supply the current for the \alfven wave and 
the deficit is compensated by the displacement current. 
In other words, the right-hand-side of linearized Ampere’s equation
\begin{equation}
   {\partial {\bf\hat B_0\cdot E_d}\over \partial t} = -4\pi {\bf\hat B_0\cdot
     j} + i c {\bf B\cdot (\hat{B}_0\times k_{aw\perp}) }, 
   \label{Efield-a}
\end{equation}
is non-zero for $n < n_c$ and that leads to a non-zero ${\bf\hat B_0\cdot 
E_d}$; $k_{aw\perp} \equiv {\bf k_{aw}}\cdot({\bf\hat B_0\times\hat B})$. 
Although, the solution of equation (\ref{Efield-a}) can be quite involved 
if the density of the medium is a non-smooth function. We will ignore that 
complication and consider the average electric field component parallel 
to the static magnetic field which can be written as
\begin{equation}
   {\bf\hat B_0\cdot E_d} =  E_d \exp(i k_{aw\parallel}\xi  - i\omega_{aw} t),
   \label{dEd}
\end{equation}
where $\xi$ is distance along a magnetic field line.

The particle motion is restricted along $\hat{\bf B}_0$, because the 
static magnetic field is strong and particles are stuck in the lowest 
Landau level. The equation for motion under the influence of the 
electric field, in the charge starvation region, in this case is
\begin{equation}
   {\d \gamma\beta\over dt} = {q E_d\over mc} \sin[\omega_{aw}(\xi/c-t)].
\end{equation}
Let us define 
\begin{equation}
   \omega_E = {q E_d\over mc}, \quad a_{aw} = {q E_d\over 
        mc\omega_{aw}}, \quad \phi = \omega_{aw}(\xi/c-t).
\end{equation}
The equation for particle motion is rewritten in terms of these as
\begin{equation}
  (1-\beta)  {\d \gamma\beta\over d\phi} = -a_{aw}\,\sin\phi \implies  
    {\d \gamma(1-\beta)\over d\phi} = -a_{aw}\,\sin\phi.
\end{equation}
The solution of this equation is
\begin{equation}
   \gamma(1-\beta) = \gamma_0(1-\beta_0) - a_{aw}(\cos\phi - \cos\phi_0),
   \label{gam}
\end{equation}
where $\phi_0$ and $\beta_0$ are the initial particle position (phase) 
and speed; $\gamma_0(1-\beta_0) \sim 1$.
If $a_{aw}$ is positive, then for $0 < \phi_0< \pi$, the particle 
acceleration is along the same direction as the \alfven wave
propagation direction, and the asymptotic value for $\cos\phi$ is
$\cos\phi_0 + \gamma_0(1 - \beta_0)/a_{aw}$. Since $a_{aw} = 2{\rm x}10^{10}
   E_{d,8} \omega_{aw,5}^{-1}$, the particle is carried by the wave and
stays very nearly at the same wave-phase. This is because
the particle is accelerated by the wave to highly relativistic
speeds in a time much shorter than the wave period, and the particle
moves with the wave at nearly a fixed phase. The difference between 
particle speed and the wave speed is $c/(2\gamma^2)$, which is extremely
small for large $\gamma$, and that is the reason that the particle
stays at roughly the initial phase angle. 

Particles of opposite charge, i.e. $a_{aw}<0$, and $0 < \phi_0< \pi$,
are accelerated along the direction opposite to the \alfven wave velocity 
vector. In the absence of radiative losses, particles are
accelerated to Lorentz factor $\sim |a_{aw}|/2$. Curvature radiation loss,
acting on particles individually, limits the particle LF to 
$\sim 10^8 E_{d,8}^{1/4} R_{B,8}^{1/2}$; where $R_B$ is the
curvature radius of magnetic field. We will see in the next 
sub-section that $\gamma\lae 10^3$ due to radiation reaction force on
particle bunches that radiate coherently.

These particles, i.e. those with $a_{aw}<0$ \& $0 < \phi_0< \pi$, 
moving close to the speed of light toward the rear end of the \alfven 
wave-packet arrive at a location in about one half wave period 
where the wave-electric-field direction has reversed, and they are then 
pulled in the opposite direction.
Their dynamics is now described by the $a_{aw}>0$ case discussed above.
These particles are trapped in a narrow region if $\gamma \ll a_{aw}$ 
and dragged with the wave for a distance $\sim c\gamma^2/\omega_{aw}$, 
which is essentially forever for the system of interest to us.

When the \alfven wave enters the charge starvation region, it drags with 
it charges of one sign in one-half-wavelength and particles of opposite
charge in the adjoining half-wavelength which has electric field of 
opposite sign. The Coulomb field due to this charge separation is 
\begin{equation}
   E_{cou} \sim 2\pi^2 q n/k_{aw\parallel}, \quad{\rm or}\quad
        E_{cou} \lae {\pi B\over 4} {k_{aw\perp}\over k_{aw\parallel}} 
\lae E_d.
\end{equation}
The inequality follows from $n < n_c$ and using equations (\ref{cur-b})
\& (\ref{n-cs-a}) for $E_d$ \& $n_c$. The Coulomb field is smaller
than the electric field ${\bf \hat B_0 \cdot E_d}$ associated with 
the \alfven wave (eq. \ref{Ed-B}).
The energy stored in the Coulomb field per unit volume is
of order, but smaller than, the \alfven wave energy density. 
Thus, when the \alfven waves enter the charge starvation region, a 
fraction of their energy gets expended to charge separation.
The kinetic energy density of particles is much smaller than the 
\alfven wave energy density ($B^2/8\pi$), since particle LF is smaller
than the maximum value of $|a_{aw}|/2$ due to the radiation reaction force. 

Electrons and positrons are advected by the 
\alfven wave at the charge starvation radius $R_c$, and the resulting
charge depletion in the interior of $R_c$ causes the starvation front to 
shift to smaller radius with time. 

Particles below $R_c$ move along $\bf\hat B_0$ at high speeds to
supply the current for the \alfven wave. Thus, there is a continuous
flux of particles crossing $R_c$ from below. These particles get 
accelerated to high LF and advected with the \alfven wave as a result 
of the strong electric field\footnote{The current density, $\bf j$, and 
the electric field $\bf E_d$ (which provides the displacement current) point 
in the same direction as per the Ampere's law. Thus, charge particles of
the sign that cross $R_c$ encounter electric field there that is pointing 
in a direction such that these particles are accelerated in the forward 
direction, i.e. to larger $R$, and not forced back to $R<R_c$. In other 
words, the electric field direction in the charge starvation region is 
such that it continuously removes charge particles from the interior.} 
in the charge starvation zone. The continuous loss of particles from 
the interior of $R_c$ causes the $e^\pm$ density to decrease with time,
and the region which was not charge starved before becomes charge
starved. We estimate the speed at which $R_c$ moves inward.

The current amplitude ($I_{aw}$) associated with the \alfven wave-packet
along ${\bf\hat B_0}$ is roughly independent of $R$; $I_{aw}$ is the
integral of current density over the cross-section of the wave-packet.  
Thus, the number of particles that cross $R_c$ per unit time, and are
subsequently advected by the \alfven wave is $\sim I_{aw}/2q$. The 
cross-sectional area of the wave packet, $A_{aw}$, increases 
as $R^3$ as the wave follows the dipole field lines. 
Therefore, $d R_c/dt \sim -I_{aw}/(2qnA_{aw})$.
This result can be recast in a slightly more convenient form as 
$d R_c/dt \sim -n_c/2n$; $I_{aw}/A_{aw} \sim j$, and $j$ can be written
in terms of $n_c$ and $n$ using equations (\ref{cur-a}) \& (\ref{cur-b}).

A corollary of this result is that $e^\pm$ present in the NS magnetosphere
are depleted (advected by the wave) on a time of order $\langle n/n_c\rangle
R_c/c$, which is a few ms.

Thus far, we have discussed how the Alfv\'en-current below $R_c$ supplies 
a flux of particles into the charge-starvation region where they are quickly
accelerated and advected by the wave. We now describe how particles
the \alfven wave-packet encounters at the head of the wave at $R>R_c$
also get trapped and advected by the wave. Charge particles of one sign --
those which experience the force of the field $\bf E_d$ in the forward 
direction -- are swept up by the wave front 
and compressed into a thin sheet of width much smaller than $\lambda_{aw}$ 
and forced to move with the wave. Particles of the opposite charge move in the
reverse direction, along $-({\bf \hat{k}_{aw}\cdot \hat{B}_0) \hat{B}_0}$, 
at speed $\approx c$ for about half a wavelength and there they start 
accumulating in a thin sheet; these results follow from equation 
(\ref{gam}) for $\gamma$. 

The Coulomb field in the thin sheet increases with time as more charges 
are swept up by the wave and accumulate in the sheet:
\begin{equation}
    E_{cou}(t) \sim \pi q n (R - R_c),   \quad\quad {\rm for}\quad R > R_c
\end{equation}
where $R$ is the location of the wavefront at time $t$. The Coulomb field 
strength exceeds the wave-field $ E_d$ for $(R-R_c)\gae \lambda_{aw}$, 
and at that point the sheet starts spreading in the longitudinal direction
at close to the speed of light. Since the
Coulomb fields at the front and the back sides of the sheet do not decrease
as the width of the sheet increases, the spreading continues until the 
sheet thickness becomes $\sim \lambda_{aw}/2$. Particles in the sheet now 
overlap with particles of opposite charge in the adjoining half-wavelength
of the \alfven packet thereby partially shielding $E_{cou}$. 

The physics of the system is that charge particles of 
one sign swept up at the front of the \alfven wave-packet, within 
$\lambda_{aw}$ of $R_c$, accumulate in a thin sheet near the 
leading front. The sheet starts to spread longitudinally, after about
one wave period, when the Coulomb field in the sheet exceeds the 
wave-electric-field. Thereafter, particles swept up at the front of 
the wave form an alternating 
positive and negative charge particle streams, separated by half a 
wavelength, that move toward the rear end of the wave-packet.
This charge particle stream partially neutralizes the charge separation 
discussed previously where the \alfven wave packet advects particles
with it from the boundary of charge starvation region at radius $R_c$.

\medskip
\noindent We close this sub-section by summarizing the main results.
\smallskip

\noindent\hangindent=10pt\hangafter=1 
1.
The \alfven wave-packet moving away from the surface of the NS becomes
   charge starved at some radius $R_c$ when the particle density falls 
   below a critical value and the plasma is unable to carry the current
   required by the wave. An electric field develops at $R>R_c$, which has a
   non-zero component along the magnetic field; at $R>R_c$ the displacement 
   current makes up for the deficit in plasma current.

\noindent\hangindent=10pt\hangafter=1 
2.
Charge particles are separated at $R_c(t)$ by the electric field, and 
    advected by different segments of the \alfven wave-packet, roughly half 
    wavelength apart, where the electric 
    force on them is in the forward direction (the direction of the wave 
    motion). The charge starvation radius moves closer to the NS with time 
    as particle density in the neighborhood of $R_c$ decreases.

\noindent\hangindent=10pt\hangafter=1 
3.
 Charge particles of one sign swept up at the head of the \alfven packet are 
    compressed into a thin sheet. When the coulomb force 
    in the sheet becomes too large to confine particles, they spread 
    longitudinally to the rear of the wave-packet. Streams of particles
 of different charges move at different speeds toward the rear of the wave, 
 and they are subject to two-stream instability that causes particles to 
 form clumps.

\medskip
 Clumps of particles advected by the \alfven wave move along the 
    curved magnetic field lines and produce coherent curvature radiation that 
    is discussed in section \ref{rad-physics}.

\subsection{Formation and acceleration of particle clumps}
\label{clump-acc}

\alfven wave packets of finite size in the direction perpendicular
to the magnetic field {\bf B}$_0$ must have a non-zero component of the
wave-vector transverse to the field. Therefore, the wave packet has a non-zero 
current along the magnetic field lines as was shown in \S\ref{alfven}. 
Electrons and positrons that carry this 
current are subject to two stream instability, and that leads to 
formation of particle clumps. The particle clumps are accelerated by the
electric field in the charge starvation region. We estimate clump LF in this
sub-section. The clumps moving along curved magnetic field lines produce
coherent curvature radiation that is responsible for the FRB signal 
(\S\ref{rad-physics}).

Growth rate of the 2-stream instability, for $e^\pm$ moving at 
sub-relativistic speeds, is obtained by solving the following equation 
(eg. Chen, 1974) 
\begin{equation}
  {\omega_p^2\over \omega^2} + {\omega_p^2 \over (\omega - k v)^2} = 1,
\end{equation}
where $v$ is the relative speed between $e^\pm$s, $(\omega, k)$ are 
frequency and wavenumber for the 2-stream instability, and 
\begin{equation}
   \omega_p = \left[{4\pi q^2 n\over m}\right]^{1/2}
  \label{wp}
\end{equation}
is the plasma frequency. The imaginary part of $\omega$ gives the growth 
rate of the instability. The growth rate for 
$k \ll \omega_p/v$ is $kv/2$. The growth rate peaks at $k\sim 1.5 \omega_p/v$
where its value is $\sim \omega_p/2$. The instability leads to formation 
of particle clumps of longitudinal size $\sim \pi v/\omega_p$.
The particle density at the charge starvation radius, $R_c$, is 
$\sim n_c$ (eq. \ref{n-cs-b}), and therefore 
the longitudinal size of a typical clump is 
\begin{equation}
  \ell_\parallel \sim c \left( {\pi m\over 4q^2 n_c}\right)^{{1\over2}}  
    \sim 0.5\, {\rm cm}\, \left[{\lambda^{aw\perp}_4 (R_{ns})\over  
    B_{11}(R_{ns})} \right]^{{1\over2}} \left[ {R_c\over 10 R_{ns}}
        \right]^{{3\over2}},
  \label{l-clump}
\end{equation}
where $\lambda^{aw\perp} \equiv 2\pi/k_{aw\perp}$, and $k_{aw\perp}$ is 
the transverse component of \alfven wave-vector.

Clumps form on the plasma time scale, which is much smaller than the 
period of the \alfven wave; the timescale for the former, for the system 
we are considering, is $\sim 0.1$ nano-second whereas the latter is longer 
than 1 $\mu$s. These results apply to
mildly relativistic velocity as well, and for the highly relativistic
motion introduction of the usual Lorentz factor gives the 
correct growth rate.

The electric field associated with \alfven waves in the charge starvation 
region is expected to be $\sim 10^8$ esu for $k_{aw\perp}/k_{aw\parallel}
\sim 0.1$ (eq. \ref{Ed-B} \& fig. \ref{fig1}), and even
in the transition region before the charge starvation radius 
the field strength is likely of order $\gae 10^5$ esu. Electrons and positrons 
are accelerated to LF of 10$^2$ when traveling a distance of 17 micro-meter 
in an 10$^8$ esu field. Were it not for the radiation reaction (RR)
forces, these particles would attain $\gamma \sim 10^{10}$
if the field extends to 1 km.
Charge particles moving along curved magnetic field lines produce
curvature radiation and the back reaction of this radiation
limits the particle LF. 
The power emitted by an electron moving with LF $\gamma$ along a magnetic
field line of curvature radius $R_B$ is 
\begin{equation}
  p_c = {2 q^2 \gamma^4 c\over 3 R_B^2}.
  \label{Pc}
\end{equation}
The characteristic frequency of curvature radiation depends on $\gamma$ \& 
$R_B$, and is given by
\begin{equation}
    \nu \approx {c \gamma^3\over 2\pi R_B}, \quad {\rm or} \quad
              \gamma \approx (2\pi \nu\, R_B/c)^{1/3} \approx 590\, 
        (R_{B,9}\,\nu_9)^{1/3}.
  \label{nu}
\end{equation}
The curvature radius of a dipole magnetic field line at location $(R,\theta)$ is
\begin{equation}
    R_B(R, \theta) \approx 0.8\, R/\theta,
   \label{Rb}
\end{equation}
where the polar angle $\theta$ is measured wrt to the magnetic axis.

The back reaction of the curvature-radiation acting
on particles individually limits their LF to $\sim 10^8 E_{d,8}^{1/4} 
R_{B,8}^{1/2}$; this is obtained by equating the rate of work done
by the electric field with the rate of loss of energy to 
radiation\footnote{This 
result is unphysical at high plasma densities
when the kinetic energy density in particle with this LF exceeds the
energy density of electric field. The electric field is shorted out
by rapid redistribution of $e^\pm$ on the plasma timescale.}.

However, radiative reaction forces do not act on particles individually when 
the wavelength of the curvature radiation is larger than the inter-particle 
distance. The calculation of collective RR force
on a particle bunch is a challenging problem. In fact, RR force calculation
on a single particle is already a hard problem. The Abraham-Lorentz
prescription for RR force for a non-relativistic particle (Landau \& 
Lifshitz, 1971),
and Dirac's generalization to particles with relativistic speed 
(Dirac, 1938) suffer from pathological properties
such as particle acceleration in advance of the force. A number
of fixes have been suggested to make RR force formulae respect
causality, however, they are ad hoc patches and not derived from first 
principles considerations.

Our best hope going forward, therefore, is to use general principles of 
energy conservation and causality to guide the estimate of Lorentz factor 
of a particle bunch that is radiating coherently and is subject to 
collective RR forces. The LF is obtained by balancing the rate of 
energy supply to a particle by the electric field and the rate of 
loss of energy to radiation as a member of a large group of particles 
radiating in phase. 

Let us consider that the electric field that accelerates particles
is turned on linearly with time, which is reasonable considering that
the displacement current $(\partial {\bf E_d}/\partial t)$ is 
compensating for the deficit in the plasma current,
\begin{equation}
    E(t) = \dot E\, t,  \quad \dot E = {E_0 \over t_0} \quad {\rm for}\;
        t< t_0.
\end{equation}
The field $E(t)$ is the electric field $E_d$ of \S\ref{c-starv} along the
worldline of a particle.
The energy equation for the charge particle before the RR force becomes 
important is
\begin{equation}
   {d\, m\gamma c^2 \over dt} = q \dot E t v,
\end{equation}
which can be rewritten as
\begin{equation}
   {d\, \gamma^2 \over dt} = {2q \dot E (\gamma\beta) t\over mc}  \quad
    {\rm or} \quad {d\, (\gamma^2-1)^{1/2} \over dt} = {q \dot E t\over mc}.
\end{equation}
The solution of this equation is
\begin{equation}
   \left[ \gamma^2(t) -1\right]^{1/2} = {q \dot E t^2\over 2mc}.
    \label{gam_Et}
\end{equation}
Or
\begin{equation}
   \gamma(t) \approx 2 (t/t_a)^2 \quad {\rm for}\quad t_a < t < t_0,
\end{equation}
where
\begin{equation}
  t_a = \left[ {12^{1/2} m c\over q \dot E}\right]^{1/2} = (4.4{\rm x}10^{-10}
  {\rm s})\; {t_{0,-7}^{1/2} \over E_{0,5}^{1/2} }.
\end{equation}

The radiation reaction force acts on a group of particles collectively that
are radiating in phase. It is convenient to define a coherent volume,
$V_{coh}$, such that 
\begin{equation}
   N_{coh} \equiv n(R)\, V_{coh} 
   \label{Vcoh-def}
\end{equation}
is the number of particles that radiate in phase, and are subject to the 
collective RR force; $n$ is the plasma density.
The coherent volume depends on $\gamma$, $\lambda$ (wavelength
of radiation), the density fluctuation spectrum of the plasma 
along the photon propagation direction [$\widetilde{\nk}(k)$] or 
typical clump size ($\ell_\parallel$), and time $t$ which sets an upper
limit on $V_{coh}$ from causality considerations.
At time $t$ -- measured from the moment when the electric field is turned on --
the coherent volume cannot be larger than $(ct)^3$. 

Curvature radiation from different electrons or positrons moving with 
LF $\gamma$ can add coherently provided that their velocity and 
acceleration vectors are nearly parallel to within an angle $\gamma^{-1}$
and they all lie inside $V_{coh}$.
If the emission region is at radius $R$, and particles are moving along
the dipole magnetic field lines, then these conditions require the transverse
 size of the coherent region, ${\cal L}_\perp$, to be no larger than 
$R/\gamma$. Combining this with the causality condition leads to
\begin{equation}
   {\cal L}_\perp \approx \min\{ct, R/\gamma, w_{aw}\},
  \label{L-perp}
\end{equation}
where $w_{aw}$ is the transverse width of the \alfven wave packet at $R$ 
or the width of the region where particles are accelerated. We will
assume for the rest of this work that $w_{aw}$ is larger than the other
two scales; it is straightforward to modify all the relevant formulae if 
${\cal L}_\perp = w_{aw}$.

The size of the coherent region in the longitudinal direction (along particle
velocity vector), ${\cal L}_\parallel$, is a more complicated entity
which depends on all the variables of the system mentioned above, viz. 
$t$, $\lambda$, $\gamma$, $\widetilde{\nk}$, $\ell_\parallel$.
The dynamical time in the rest frame of a clump at radius $R$ that is moving 
with LF $\gamma$ is $R/c\gamma$, and the size of the region in causal contact
in this frame is $R/\gamma$. Therefore, the maximum size of causally 
connected region in the lab frame is $R/\gamma^2$. Let us consider that
a clump has been accelerated to $\gamma$ by an electric field turned on 
time $t$ ago.
If $ct$ is smaller than the characteristic wavelength of curvature 
photons\footnote{Technically it is incorrect to talk about $\lambda > ct$, 
as photons of wavelength $\lambda$ can not be produced in time less than 
$\lambda/c$. However, what we are quantifying here is the longitudinal
size of the region where all particles (of the same charge) are radiating
in phase, and that size cannot exceed $ct$, i.e. ${\cal L}_\parallel\lae
  ct$.}, then causality suggests that ${\cal L}_\parallel \sim \min\{ct,
R/\gamma^2\}$. 

Curvature radiation from particles separated by $\lambda/2$ along the 
photon propagation direction nearly cancels when $ct \gg \lambda$.
The contribution to the luminosity from a region of longitudinal
size $\lambda$ is non-zero when 
$\widetilde{\nk}(2\pi/\lambda) \not= 0$.
Let us consider that the medium is clumpy and the typical clump size is
$\ell_\parallel$ in the longitudinal direction.
For $\lambda > \ell_\parallel$, the net emission from a region of size
$\lambda$ depends on the difference between the number of particles in the
adjacent two half-wavelengths, which is roughly proportional to 
$(\lambda/\ell_\parallel)^{1/2}$. Therefore, ${\cal L}_\parallel\sim 
\ell_\parallel$ in this case as different particle clumps are randomly 
placed within a wavelength and the electric fields of radiation from them
have random relative phases.
Similar considerations for $\lambda < \ell_\parallel$ lead to 
${\cal L}_\parallel\sim |\widetilde{\nk}(k)|/n$ with 
$k\sim \pi/\lambda$. Results for these different cases are summarized
in the following equation
\begin{equation}
   {\cal L}_\parallel\sim \left\{ 
    \begin{array}{l}
     \hskip -5pt  \min\{ct, R/\gamma^2\}  \hskip 45pt   ct < \lambda\; \& \;
     \ell_\parallel \\ \\
   \hskip -5pt   \ell_\parallel  \hskip 94 pt  
     ct > \lambda > \ell_\parallel \\ \\
   \hskip -5pt  \bigl|\widetilde{\nk}\bigl(\pi/\lambda\bigr)\bigr|/n  
    \hskip 57 pt  ct > \lambda < \ell_\parallel 
    \end{array}
  \right.
   \label{L-par}
\end{equation}

The number of particles that experience collective radiation-reaction
force can be rewritten in terms of ${\cal L}_\parallel$ \& 
${\cal L}_\perp$ as
\begin{equation}
   N_{coh} = n\, V_{coh} \sim n\, {\cal L}_\parallel {\cal L}_\perp^2.
   \label{Ncoh}
\end{equation}
The acceleration of a clump passes through a few different stages of 
the nine cases covered by equations (\ref{L-perp}) and (\ref{L-par}) 
depending on various parameters of the system.

Initially, the coherent volume grows with time as, $V_{coh} \sim n(ct)^3$ 
[eqs. \ref{L-perp}) \& (\ref{L-par}].
Balancing the rate of work done by the electric field on particles 
within $V_{coh}$ with the rate of loss of energy to curvature radiation 
by these particles collectively, we find
\begin{equation}
   q E(t) c \approx p_c N_{coh} \implies \gamma \approx \left[ { 3 \dot E
       R_B^2 \over 2 q n c^3 t^2}\right]^{1/4}.
    \label{gam_t3}
\end{equation}
The time $t_{rr}$ when the RR force becomes important is when $\gamma$ 
given by equations (\ref{gam_Et}) and (\ref{gam_t3}) are roughly equal, i.e.
\begin{equation}
   {q \dot E t^2\over 2mc} \sim \left[ { 3 \dot E
       R_B^2 \over 2 q n c^3 t^2}\right]^{1\over4},
\end{equation}
or
\begin{equation}
    t_{rr} \sim \left[ {24 R_B^2 m^4 c\over q^5 {\dot E}^3 n}\right]^{1\over10} 
     \sim (5{\rm x}10^{-9}{\rm s}) R_{B,8}^{1\over5} {\dot E}_{12}^{-3\over10} 
    \n13^{-1\over10}.
\end{equation}
The LF for $t < t_{rr}$ is given by equation (\ref{gam_Et}) and for 
$t > t_{rr}$ by equation (\ref{gam_t3}). The particle LF and the wavelength 
of the curvature radiation at time $t_{rr}$ are
\begin{equation}
   \gamma(t_{rr}) \sim 300\, (R_{B,8}{\dot E}_{12})^{{2\over 5}} 
     \n13^{-{1\over 5}}, \quad \lambda(t_{rr}) \sim (20\,{\rm cm}) 
  R_{B,8}^{-{1\over 5}} {\dot E}_{12}^{-{6\over 5}} \n13^{{3\over 5}}.
\end{equation}
If the parameters are such that $ct_{rr} < \lambda$, then with time this 
inequality grows only bigger (as $\lambda\propto \gamma^{-3}\propto t^{3/2}$)
and the evolution of $\gamma$ continues to be described by equation 
(\ref{gam_t3}) for a while. 
However, for the parameters relevant for FRBs, $ct_{rr} > \lambda$, $\lambda 
\gae \ell_\parallel$, and $c t_{rr} < R/\gamma$. Therefore, the coherent volume 
for the calculation of RR force should be $V_{coh} \sim \ell_\parallel 
(ct)^2$ instead of $(ct)^3$. The evolution of clump LF in this case
is provided by the following equation:
\begin{multline}
   q \dot E\,t\, c \approx p_c n V_{coh} \implies \gamma(t) \approx 
      \left[ {3 {\dot E} R_B^2 \over (2 q n c^2 t
    \ell_\parallel) }\right]^{{1\over4}}, \\ 
    \hskip 55 pt {\rm or} \quad \gamma(t) \sim 1.4{\rm x}10^3 \left[ 
    { {\dot E}_{12} R_{B,9}^2 \over \n13 t_{_{-7}}\ell_\parallel}
        \right]^{1\over4}.\quad\quad
    \label{gam_t2}
\end{multline}
The time when the collective RR force becomes important and starts to dominate
the clump dynamics is obtained from the following equation
\begin{multline}
   {q \dot E t^2\over 2mc} \sim \left[{3 {\dot E} R_B^2 \over 2 q n c^2 t
    \ell_\parallel}\right]^{{1\over4}} 
  \implies t_{rr} \sim \left[ { 24 c^2\, m^4 R_B^2\over q^5 n {\dot E}^3
       \ell_\parallel}\right]^{{1\over9}}, \\
   \hskip 25 pt {\rm or}\quad t_{rr} \sim (9{\rm x}10^{-9}{\rm s})\, 
     R_{B,8}^{2\over9}\n13^{-{1\over9}} {\dot E}_{12}^{-{1\over3}}
            \ell_\parallel^{-{1\over9}}. \quad\quad\quad
   \label{trr2}
\end{multline}
The LF decreases with time as $t^{-{1\over4}}$ for $t>t_{rr}$ if the 
electric field increases linearly with time; the decrease is slower if 
$\dot E$ rises with time, and for $t<t_{rr}$ the LF is given approximately
by equation (\ref{gam_Et}).
The LF of the particle clump at $t_{rr}$ is 
\begin{equation}
    \gamma(t_{rr}) \sim {q \dot E t_{rr}^2\over 2mc} \sim 700
         { {\dot E}_{12}^{{1\over3}} R_{B,8}^{4\over9} \over (\n13
            \ell_\parallel)^{2\over9} },
\label{gam-trr}
\end{equation}
and the wavelength of curvature photons at this time is
\begin{equation}
   \lambda(t_{rr}) \approx {2\pi R_B\over \gamma^3(t_{rr})} \sim
      (1.1\,{\rm cm}) {\dot E}_{12}^{-1} R_{B,8}^{-{1\over3}} (\n13
        \ell_\parallel)^{2\over3}.
    \label{lambda-trr}
\end{equation}
Thus, $\lambda(t_{rr}) < c t_{rr}$, and the clump LF was calculated with 
$V_{coh}$ in the correct regime (eq. \ref{L-par}). 
Since, $\gamma\propto ({\dot E}/t)^{{1\over4}}$ and $\lambda \propto 
\gamma^{-3} \propto (t/{\dot E})^{{3\over4}}$, for the entire visible 
worldline of the clump $\lambda/ct < 1$.

At the time, $t_{rr}$, when the RR force begins to dominate the dynamics, 
$R/\gamma > ct_{rr}$ (for $R\gae 10^6$cm, see eq. \ref{trr2}) and this 
ordering is preserved up to time $\sim (3\,\mu{\rm s}) R_7^{4\over3}
(\n13\ell_\parallel)^{1\over3}/({\dot E}_{12}^{1\over3} R_{B,8}^{2\over3})$.
The radiation we observe from a particle clump is produced
over a period of time of order $\sim R_B/(\gamma c)\sim 10^{-4}$s 
in the lab frame. Therefore, the clump acceleration switches to 
the regime where ${\cal L}_\perp \sim R/\gamma$ after a few $\mu$s,
and the coherent volume becomes $V_{coh} \sim
  \ell_\parallel (R/\gamma)^2$. This 
leads to the following equation for the clump LF:
\begin{equation}
   \gamma \sim {R_B\over R}\left[ {3 E(t)\over 2q n\ell_\parallel}
    \right]^{1\over2} \sim 300 {R_{B,8}\over R_7} \left[ {E_{7} \over 
   \ell_\parallel\n13} \right]^{1\over2},
   \label{clump-LF1}
\end{equation}
which is obtained, as before, by balancing the energy gain and loss rate for 
particles in the clump. Since the particle density in clumps should be close
to the critical density given by equation \ref{n-cs-b}, the above equation
can be rewritten as
\begin{equation}
   \gamma \sim 800 \left[ {E_7(R)\over \ell_\parallel B_9(R)} \right]^{1\over2}
       { [\lambda^{aw\perp}_{,4}(R_{ns})]^{1\over2}
        R_{B,8}\over R_7^{1/4} },
   \label{clump-LF2}
\end{equation}
where $E$ and $B$ are electric and magnetic field perturbation amplitudes
associated with the \alfven wave packet at radius $R$. The clump LF has
a weak dependence on the radius where the wave becomes charge starved, 
although it might not seem that way because of the factor $R_B$ in the
numerator. The clump size $\ell_\parallel$ is proportional to plasma
length scale ($\sim c/\omega_p$) which is $\propto R^{3/2}$, and 
$R_B\propto R$, hence $R_B/(R^{1/4} \ell_\parallel^{1/2})$ is independent 
of radius, and that means that the clump LF is nearly $R$-independent.

An important point to note is that the electric field must be turned 
on gradually -- to be quantified shortly -- to ensure that the volume of 
the coherent region is large and the clump LF plateaus at $\sim 10^3$
instead of increasing uncontrollably to $\gamma\sim 10^8$, which is the
limit when RR force acts on particles individually.
The characteristic wavelength of curvature photons at $t_{rr}$ is 
given by equation (\ref{lambda-trr}). If this wavelength is smaller than
$\sim \ell_\parallel$ then the coherent volume decreases as 
$\widetilde{\nk}(\pi/\lambda)$ (eq. \ref{L-par}). The smallest clump 
size is expected to be not much smaller than the plasma length scale 
according to the two-stream instability (discussed at the beginning
of this sub-section). In that case, $\widetilde{\nk}$ likely 
decreases exponentially for $k \gae \ell_\parallel^{-1}$.
So, if $\lambda(t_{rr})$ were to be smaller than $\ell_\parallel$, 
then $\gamma$ would increase with time uncontrollably -- an increase 
of $\gamma$ produces photons of smaller wavelength and that shrinks 
the coherent volume, which then reduces RR force causing 
$\gamma$ to increase further. The unstable growth of $\gamma$ is avoided 
only when $\lambda(t_{rr}) > \ell_\parallel$, and the electric 
field turns on slowly such that
(see eq. \ref{lambda-trr})
\begin{equation}
   \dot E \lae (10^{12}\,{\rm esu\, s}^{-1}) \n13^{2\over3}/(R_{B,8}
       \ell_\parallel)^{1\over3}.
\end{equation}
As long as the electric field ramps up on the time
scale of the \alfven wave period, $\sim 10 \mu$s, particles within 
a clump can remain in causal contact while they are being accelerated 
by the electric field. They are then subjected to the large collective,
radiation reaction force, and $\gamma$ evolves stably. It should be 
pointed out that clumps moving in the same direction as the \alfven wave,
do not ``see'' the electric field increase with time since they ride the 
wave at essentially a fixed phase when their $\gamma\gae 10^2$.
It follows from the above equation that $E(t_{rr}) \lae 10^4$esu, since
$t_{rr}\sim 10^{-8}$s (eq. \ref{trr2}).

We have investigated the evolution of particle clump LF for several 
different parameter regimes. The LF can be related to the effective
volume of the region ($V_{coh}$) within which particles suffer the 
collective RR force. The coherent volume can be considered to be primarily 
a function of $\gamma$ for describing the clump dynamics; other parameters,
such as plasma density fluctuations and magnetic field curvature radius, 
are independent of the clump motion to the lowest order.
If $V_{coh}(\gamma)$ were to decline faster than $\gamma^{-4}$ 
then the rate of loss of energy per particle, which is 
proportional to $p_c N_{coh}\propto \gamma^4 N_{coh}$, decreases with 
increasing $\gamma$. This is an unstable situation; as $\gamma$
increases, particles lose less energy and thereby LF goes up.
The runaway increase of $\gamma$ is terminated at a very large value
when the electric field energy is depleted or the field shorted by charge 
redistribution, or the rate of energy loss for an individual particle
to curvature radiation is balanced by the rate of work done by the electric
field.

We close this sub-section by pointing out the obvious, viz. that the 
calculations presented here for the Lorentz factor of particle clumps 
have a higher degree of uncertainty than other parts of the paper. 
The LF is determined 
by balancing the electric field acceleration of particles in the clump 
with the collective radiation reaction force. The latter is not possible 
to calculate from first principles. We have, instead, relied on energy 
conservation and causality arguments to provide a rough estimate for the 
RR force and estimate the clump LF for various possibilities. 

\section{Coherent Curvature Radiation}
\label{rad-physics}

Electrons and positrons move in opposite directions along the
magnetic field at speed $\sim c$ below the charge starvation radius, $R_c$, 
to carry the current required by the \alfven wave-packet. 
Moreover, electrons and positrons swept up at the head of the \alfven wave
packet at $R>R_c$ move at different speeds as the electric forces on $e^+$
and $e^-$ are in opposite directions. 
The differential motion between $e^\pm$ is subject to the two-stream 
instability which leads to formation of clumps as discussed at the
beginning of \S\ref{clump-acc}. The fastest growing modes grow at a 
rate $\omega_p$, and the longitudinal size of the clumps that form due 
to 2-stream instability, $\ell_\parallel$,  is of order the plasma length 
scale, $\sim \pi c/\omega_p$, which is given by equation (\ref{l-clump}). 

The curvature radiation produced by electrons moving with LF $\gamma$,
in a region of transverse size $\sim R/\gamma$, adds up 
coherently\footnote{The angle between magnetic field lines,
for a dipole magnet, at two locations at the same radius but separated by
an angle $\delta\theta$ can be shown to be
$3\delta\theta(1 +\cos^2\theta)/(1 + 3\cos^2\theta)$ which is
$3\delta\theta/2$ near the pole, and $3\delta\theta$ at the
equator.}$^,$\footnote{Consider
two photons produced at the same time in the lab frame via the
curvature radiation mechanism. One of these photons is emitted by
an electron located at $(R, \theta=0)$ and the other one at
$(R+R/2\gamma^2, \theta=\gamma^{-1})$ arrive at the observer at the
same time. The magnetic field directions
at these two locations are within an angle $\gamma^{-1}$ and so are the
velocity and acceleration vectors of particles at these places.
Therefore, both these photons arrive at the observer at the same time
with the same phase angle, and hence add up coherently.}, 
since all these electrons have the same acceleration vectors to 
within an angle $\gamma^{-1}$ and their velocity vectors are also parallel to 
within $\gamma^{-1}$.

Consider two particles located along the line of sight to the observer
and separated by distance $\delta r$. These particles are
moving with LF $\gamma$ toward the observer. The relative speed between 
photons and these particles in lab frame is $c/2\gamma^2$. Therefore, 
the particle in the back should emit a photon before the particle in 
the front by an amount of time $\delta t\sim 2\gamma^2\delta r/c$ 
(in lab frame) so that these photons arrive at the observer at the 
same time. If the particle in the front is at a distance $R$ from the
compact object then it has been moving for a time $\lae R/c$ in the 
lab frame. Since $\delta t < R/c$, this construct shows that the
largest causally connected radial distance from which photons can arrive 
at the observer at the same time is $R/\gamma^2 \equiv \Delta R_{cau}$.

Particle clumps attain a roughly constant LF $\sim 10^3$ in a short time 
of a few $\mu$s as a result of the electric field acceleration and 
radiation reaction drag force acting collectively on particles in the
clump (see \S\ref{clump-acc}).
During the coasting phase, the total number of particles that radiate 
in phase is (see eqs. \ref{L-perp} \& \ref{L-par})
\begin{equation}
    N_{coh} \sim n \ell_\parallel (R/\gamma)^2,
   \label{Nc}
\end{equation}
and the total number of particles from which an observer receives 
photons at a fixed time is larger than $N_{coh}$ by a factor 
$\sim \Delta R_{cau}/\ell_\parallel = R/(\ell_\parallel\gamma^2)$. 
Making use of equation (\ref{Pc}) for curvature power per electron ($p_c$) 
and equation (\ref{Nc}) we obtain the total curvature power to be
\begin{equation}
  P_{c} = N_{coh}^2(\Delta R_{cau}/\ell_\parallel) p_c = {2 q^2 c R^5 
    \ell_\parallel n^2 \over 3 R_B^2 \gamma^2}.
  \label{p-clump}
\end{equation}
The luminosity in the observer frame is
\begin{equation}
   L_{iso} \approx 8 \gamma^4 P_{c} \approx
   {16 q^2 c R^5 \gamma^2 n^2 \ell_\parallel \over 3 R_B^2}.
   \label{Liso-a}
\end{equation}
We can use equation (\ref{nu}) to replace $\gamma$ in favor of $\nu$ and 
express the FRB luminosity in a more convenient form
\begin{equation}
   L_{iso} \approx {16 (2\pi)^{2/3} q^2 c^{1/3} R^5 n^2 \ell_\parallel 
    \nu^{2/3} \over 3 R_B^{4/3} }.
\end{equation}
Replacing $\ell_\parallel$ using equation (\ref{l-clump}), and $n\approx
n_c$ (eq. \ref{n-cs-b} gives $n_c$), leads to
the following expression for the luminosity
\begin{multline}
  L_{iso}\approx (10^{44} {\rm erg\, s^{-1}}) \left[ {B_{11} \over 
        \lambda^{aw\perp}_{,4}}\right]^{{3\over2}} {\nu_9^{{2\over3}}\over 
     R_{B,8}^{{4\over3}} }\left[{R_c\over R}\right] 
  \left[ {R_c\over 10 R_{ns}}\right]^{{1\over2}},
   \label{Liso}
\end{multline}
where $R_c$ is the charge starvation radius, $R$ is the radius where 
curvature radiation is produced (which should be the same as $R_c$ or 
very close to it), $\lambda^{aw\perp} \equiv 2\pi/k^{aw\perp}$, 
$k^{aw\perp}$ is the transverse component of \alfven wave-vector
and $B$ is \alfven wave amplitude, both of which are evaluated at the 
wave launching site, i.e. at the NS surface; $k^{aw\perp}$ is larger 
than $k^{aw\parallel}=\omega_{aw}/c$ at the surface of the NS by a
factor $\sim$10--10$^2$. The magnetic field produced by the motion of 
charge clumps must be smaller than the original field by at least
a factor $\gamma$ in order to maintain coherent addition of radiation 
from different particles in the clump (Kumar et al. 2017). This
condition is satisfied at $R_c$ provided that the magnetic field 
strength at the NS surface is $\gae 10^{14}$G.

The specific luminosity $L_{iso}^\nu \sim L_{iso}/\nu\aprop \nu^{-1/3}$.
The dependence of luminosity on frequency is more complicated than 
suggested by this equation; $L_{iso}^\nu$ depends on the distribution of 
clump sizes and their LFs. In terms of the Fourier transform of particle 
number density $n({\bf x})$, i.e. ${\widetilde\nk}(k)$, $L_{iso}^\nu 
\aprop \nu^{2/3} |\widetilde{\nk}({\bf k})|^2$ in the high frequency limit
of $\nu \gae c/\ell_\parallel$ (see eq. \ref{L-par}); 
where $|{\bf k}| = 2\pi\nu/c$.

If the minimum clump size is the plasma length scale or $\sim \ell_\parallel$,
 as suggested by the two-stream instability, then $|\widetilde{\nk}(k)|$
falls off exponentially for $\nu \gae c/\ell_\parallel$ ($\ell_\parallel$ is 
given by eq. \ref{l-clump}). Therefore, the luminosity is expected fall off
sharply at $\nu \gae c/\ell_\parallel$. 

The electric field develops before the \alfven wave reaches the 
charge starvation radius $R_c$ as the plasma is clumpy due to 2-stream
instability and possibly other effects. However, the field strength at 
$R< R_c$ is smaller than $E_d$ (eq. \ref{Ed-B}), which is 
the value of the electric field in the fully charge starved region. 
Let us take the field strength in the transition zone to be $\eta E_d$; 
$\eta$ is a function of radius $R$. The electric force is balanced by 
the radiation reaction force acting collectively on particles in the clump 
as discussed in \S\ref{clump-acc}. In this case, the rate of energy radiated 
per unit volume is equal to the work done by the electric field $E$, i.e. 
\begin{equation}
    \epsilon_{em} = q n c\, \eta E_d \sim q n c\, \eta B 
     (k_{aw\perp}/k_{aw\parallel}).
\end{equation}
The energy density in \alfven waves is $B^2/4\pi$, and therefore the
dissipation length for the wave is
\begin{equation}
    d_{aw} = {c B^2 \over 4\pi \epsilon_{em}} \sim {B \over 4\pi q n \eta} 
          \left[ { k_{aw\parallel} \over k_{aw\perp} }\right] \sim 
    {\rm 2 x10^5\, cm}\, {B_{10}(R)\over \eta\,\n13},
  \label{daw}
\end{equation}
where $B/n \propto R^{3/2}$ is calculated at the charge starvation radius. 
The \alfven wave damps rapidly as it approaches the charge starvation
radius; the dissipation length is roughly of order $\eta^{-1}$ times the 
wavelength of the waves, and the wave loses a large fraction of its energy
in the vicinity of $R_c$. 

\subsection{Predictions of the model}

\noindent\hangindent=9pt\hangafter=1
$\bullet$ {\bf Maximum frequency for FRBs}

\noindent\hangindent=9pt\hangafter=0
The EM spectrum is cutoff exponentially for wavelengths much smaller
than particle clump size or the longitudinal size of the coherent region. 
We find the cutoff frequency using equation (\ref{l-clump}):
\begin{equation}
  \nu_{max}\sim c/\ell_\parallel \sim 50{\rm GHz}\,\left[{ B_{11}(R_{ns}) \over 
     \lambda^{aw\perp}_4(R_{ns})} \right]^{1/2} \left[ {10 R_{ns}\over R_c}
      \right]^{3/2}.
\end{equation}
This assumes that clumps have a distribution of $\gamma$, which they do when 
the ratio of the electric field strength (the component 
parallel to the static magnetic field) and the wave-magnetic field 
($E/B$) fluctuates, perhaps as a result of large density 
variation near the charge starvation region (eqs. \ref{clump-LF1} \&
\ref{clump-LF2}). The $\gamma$-distribution can 
only be determined by a very demanding numerical simulations of \alfven
waves in charge starvation region that includes the 2-stream instability.
Coherent radiation at a frequency much larger than 50 GHz can be produced 
only if plasma density fluctuations were to extend well below the plasma 
length scale ($\sim c/\omega_p$), which is unlikely but cannot be ruled out 
\begin{equation}
 \nu \propto \gamma^3/R_B \aprop
 \ell_\parallel^{-3/2} (R_B^2/R) (E/B)^{3/2} \lambda_{aw\perp}^{3/2}
\end{equation}

\smallskip
\noindent\hangindent=9pt\hangafter=1
$\bullet$ {\bf Minimum frequency for FRBs} 

\noindent\hangindent=9pt\hangafter=0
The maximum wavelength of radiation for particle clumps moving with LF $\gamma$
is given by the radial size of causally connected region, i.e. $R/2\gamma^2$,
which is larger than the typical $\lambda$ of radiation from these bunches
by a factor $\gamma R/(2\pi R_B)\sim 20\, \gamma_3 R_7/R_{B,8}$; $R$ is the 
radius at which radiation is produced and $R_B$ is the radius of curvature 
of magnetic field lines there. Equating the curvature radiation wavelength
($2\pi R_B/\gamma^3$) with $R/2\gamma^2$, we find the LF of clumps which
produce radiation of characteristic wavelength the size of causally 
connected region
\begin{equation}
   \gamma\sim {4\pi R_B\over R_c} \implies \lambda_{max} \sim {2\pi 
     R_B\over\gamma^3}  \sim 300\;{\rm cm}
        {R_{c,7}^3\over R_{B,8}^2};
\end{equation}
where $R_c$, as before, is the charge starvation radius for the \alfven wave
where the FRB radiation is produced, and $R_B$ is the radius of curvature
of magnetic field lines in the radiation region at $R_c$.
So, the minimum frequency of radiation according to the model is $\nu_{min}\sim
10^2$MHz $R_{B,8}^2/R_7^3$. The size of clumps which have the appropriate LF
for producing $\nu_{min}$ can be obtained from equation (\ref{clump-LF1}) and
is given by $\sim 40$ cm $R_7^{3/2} E_7 \lambda_4^{aw\perp}/B_9(R)$.

\smallskip \noindent\hangindent=9pt\hangafter=0
The transverse size of the region from which an observer receives
photons is the smaller of the length scales $\sim R/\gamma\aprop R/\nu^{1/3}$
and the transverse size of the \alfven wave-packet where the
radiation is produced ($w_{aw\perp}$). If $w_{aw\perp} < 2\lambda_{max}\gamma$
then the observed luminosity declines as $\nu^{4/3}$ at low frequencies 
modulo the power spectrum of fluctuations in $\gamma$ on the
corresponding length scale.

\smallskip
\noindent\hangindent=9pt\hangafter=0
The discussion, thus far, has been about the intrinsic source properties. 
However, as the radiation propagates through the circum-stellar medium,
it is subjected to strong induced-Compton scatterings and loss of 
energy as particles along its path are accelerated to large Lorentz factors 
out to distances of $\sim 10^{14}$cm of the source (Kumar and Lu, 2019). 
The magnitude of these propagation 
effects increases with decreasing photon frequency roughly as $L_\nu/\nu^2$,
which can become substantial at $\lambda_{min}$ depending on the 
unknown properties of the CSM of FRBs.

\smallskip
\noindent\hangindent=9pt\hangafter=1
$\bullet$ {\bf Minimum \& maximum luminosities for FRBs}

\noindent\hangindent=9pt\hangafter=0
Low luminosity \alfven waves might never become charge starved, 
and hence, according to the model presented here, there should be a minimum
floor for the FRB luminosities. The absolute minimum is set by the
consideration that the charge starvation density ($n_c$) is equal to the
Goldreich-Julian density ($\nGJ$). Since plasma density in NS magnetosphere
is $\gae \nGJ$, when $n_c < \nGJ$ the \alfven wave packet can travel 
 without becoming charge starved to large distances from  NS surface.
Therefore, particles clumps don't get accelerated to produce coherent 
radiation. The condition that $n_c\gae \nGJ$
implies that the \alfven wave amplitude
$B(R_{ns}) \gae 7{\rm x}10^{8} B_{ns,15} \lambda^{aw\perp}_{,4}/P_{ns}$Gauss
 (eqs. \ref{n-cs-b} \& \ref{nR}); where $B_{ns}$ is NS surface magnetic field,
$P_{ns}$ is its rotation period, and $\lambda^{aw\perp}$ is the transverse
cross-section of the \alfven wave packet at the wave launching radius. Thus,
the \alfven wave luminosity at the NS surface, before correcting
for the finite beam angle, is $L_{aw} \sim 10^{36}
[B_{ns,15} \lambda^{aw\perp}_{,4}]^2/P_{ns}^2$erg s$^{-1}$. The isotropic
luminosity of the wave at the radius where it becomes charge starved 
and produces FRB radiation is $L_{aw}^{iso}(R_c) \sim B(R_{ns})^2 [R_{ns}/R_c]^3
R_c^2\sim 10^{39} [B_{ns,15} \lambda^{aw\perp}_{,4}]^2/(R_{c,7}\,P_{ns}^2)$ 
erg s$^{-1}$. The efficiency for converting \alfven wave luminosity to FRB
radiation is very high as discussed below equation (\ref{daw}), therefore,
the minimum FRB luminosity is 
\begin{equation}
  L_{FRB}^{min} \sim 10^{39}{\rm erg\, s^{-1}} { \left[B_{ns,15} 
      \lambda^{aw\perp}_{,4}\right]^2 \over R_{c,7}\,P_{ns}^2}.
\end{equation}
The $L_{FRB}^{min}$ calculated above is for an observer whose line
of light lies within the relativistic cone of the source. The bolometric 
luminosity outside the relativistic beam falls off rapidly with increasing 
angle ($\theta$) from the edge of the cone as $(\theta\gamma)^{-6}$, and 
the peak of the spectrum shifts to lower frequencies as $(\theta\gamma)^{-2}$.

\noindent\hangindent=9pt\hangafter=0
The maximum luminosity for FRBs is estimated to be $\sim 10^{49} R^2_{B,8}
B_{ns,15}$ erg s$^{-1}$ as discussed in Lu and Kumar (2019).

\smallskip
\noindent\hangindent=9pt\hangafter=1
$\bullet$ {\bf Beaming angle and true energy}

\noindent\hangindent=9pt\hangafter=0
There are three different angular scales in the system. The first one
is the angular size of the region about the magnetic pole from which FRB
producing \alfven waves are launched ($\theta_{aw}$). 
The other two are the relativistic beaming, and the angular size of the 
FRB source. 

\smallskip
\noindent\hangindent=9pt\hangafter=0
In order for an \alfven wave packet to produce an FRB, it 
should be able to travel to large enough distances from the neutron star
surface where it becomes charge starved. 
A magnetic field line emerging from the NS surface at polar angle 
$\theta$ wrt to the magnetic pole reaches a maximum distance 
of $R_{ns}/\sin^2\theta$ and then curves back toward the NS surface.
Hence, \alfven waves that originate within an angle $\theta_{aw} \lae 0.2$ 
rad of the magnetic pole can travel out to a few 10s of $R_{ns}$, and have a 
greater chance of running into a low density plasma, become charge starved, 
and produce a FRB. 

\smallskip
\noindent\hangindent=9pt\hangafter=0
The LF of particle clumps that produce GHz photons via curvature radiation 
is of order 10$^3$ (eq. \ref{nu}), and these photons are beamed in a cone 
of opening angle $\sim 10^{-3}$rad. However, a burst is observable over a 
much larger fraction of the sky, which is the angular size of \alfven wave 
packet at the radius where the coherent radiation is produced. \alfven wave 
packets move along diverging dipole magnetic field lines and their transverse
size increases with radius as $\lambda^{aw\perp}(R_{ns}) [R/R_{ns}]^{3/2}$. 
Thus, the angular size of the wave packet at radius $R$ is 
\begin{equation}
   \theta_{frb} \sim {\lambda^{aw\perp}(R_{ns})\over R_{ns}}
    \left[{ R\over R_{ns} }\right]^{1\over2} 
    \sim 2{\rm x10^{-2}\, rad}\; \lambda^{aw\perp}_{,4}\, R_7^{1\over2}.
\end{equation}
The total energy associated with a FRB event that has isotropic luminosity
$10^{44}$erg s$^{-1}$ and duration 1 ms, after correcting for the
beaming, is 
\begin{equation}
  E_{frb} \sim {L_{frb} t_{frb} \theta_{frb}^2\over4} \sim
    10^{37}{\rm erg} R_7 [\lambda^{aw\perp}_{,4}(R_{ns})]^2.
\end{equation}

\noindent\hangindent=9pt\hangafter=0
The probability of seeing a burst from a random magnetar is $\theta_{frb}^2/4
\sim 10^{-4}$. If a magnetar produces a larger number of bursts, 
we expect them all to be within a solid angle $\pi\theta^2_{aw} (R/R_{ns})$
based on the above argument\footnote{The fraction of the sky where one
of the bursts from the object can be seen is larger than 
$\pi\theta^2_{aw} (R/R_{ns})$ if the angle between the magnetic and 
rotation axis is $\gae \theta_{aw} (R_c/R_{ns})^{1/2}\sim 1$rad. 
This is because, in this case, the sweep of the rotation axis enlarges 
the solid angle coverage of \alfven wave packets.}.
So the probability for a random observer to see one of these large number 
of bursts is $\theta^2_{aw} (R/R_{ns})/4\lae$10\%. The probability for an FRB
from one of the $\sim 30$ magnetars in our galaxy might be considerably 
smaller if the FRB activity is associated with young magnetars; magnetars 
in our Galaxy are $\gae 10^3$ years old (Tendulkar, 2014).

\smallskip
\noindent\hangindent=9pt\hangafter=1
$\bullet$ {\bf Periodicity of repeated bursts from an object?}

\noindent\hangindent=9pt\hangafter=0
According to our model, only those \alfven waves that are launched 
from the NS surface within a distance $\theta_{aw}R_{ns}$ of the magnetic 
pole can travel out to a distance of a few 10s of NS radii, where they become 
charge starved and produce FRBs; $\theta_{aw} \sim 0.2$ as discussed under
the previous bullet point. The angular size of an \alfven wave packet moving
along magnetic field lines increases with radius as $R^{1/2}$. Thus,
we expect coherent radio waves to be produced within a cone of 
angular size $\theta_{aw}(R_c)\sim \theta_{aw} (R_c/R_{ns})^{1/2}\sim 1$ rad, 
around the magnetic axis, and occurring almost randomly in time. 
If the angle between the rotation and the magnetic axis is much smaller 
than $\theta_{aw}(R_c)$ then there is no preferred phase of the NS rotation 
when we see bursts. However, if the inclination between the two axis is 
of order $\theta_{aw}(R_c)$, or larger, then we would only see 
bursts from an object during a limited NS rotation phase when the 
\alfven cone is pointing at us. This effect is somewhat more pronounced for 
stronger bursts, which are produced by \alfven waves of higher luminosity,
which become charge starved at a smaller radius and have smaller 
$\theta_{aw}(R_c)$. This is something that observers can investigate 
when we have data for a large number of bursts from an object such 
as FRB 121102.

\section{Discussion}

We have described a model for FRBs where large amplitude 
\alfven waves are produced by some disturbance at the surface of a 
magnetar in the magnetic pole region, and as these waves travel to larger
radii they decay and produce coherent radiation (Fig. \ref{fig-scenario}
shows the main components of the model). \alfven wave 
packets require a non-zero current along the static magnetic field as long 
as the transverse component of their wave-vector is non-zero. The current 
becomes stronger as the wave travels to larger radii. The counter-streaming 
$e^\pm$s that carry the current are subject to the two two-stream instability, 
which leads to formation of particle bunches of size of order the plasma 
scale. At some large radius, of order a 10 $R_{ns}$, $e^\pm$ density is 
insufficient to supply the current the wave packet requires. This is 
the region of charge starvation for the \alfven wave.
Strong electric field develops along the magnetic field in the 
charge starvation region, which provides the 
displacement current to make up for the deficit in $e^\pm$ current. 
The electric field sweeps up particles and accelerates them to different
speeds depending on their charges. The resulting differential motion
also leads to clump formation. Particle bunches are accelerated by the
electric field to LF $\sim 10^3$. They follow the curved magnetic
field lines and produce the powerful coherent FRB radiation. 

The \alfven wave luminosity at the launching site is $B^2 c 
\lambda_{aw\perp}^2/4\pi$; where $\lambda_{aw\perp}$ is the transverse size
of the wave packet. The transverse size of the wave packet increases with
radius as $R^{3/2}$ as the wave follows the diverging dipole magnetic 
field lines of the magnetar. Thus, the 
wave packet cross-section area at the radius $R$ where the radiation 
is produced is $\lambda_{aw\perp}^2 (R/R_{ns})^{3}$, and the angular size of
the beam over which the radiation is spread is $(\lambda_{aw\perp}/R_{ns})
 (R/R_{ns})^{1/2}$. 
The efficiency for conversion of \alfven wave luminosity to radio waves 
is very high in our model. Hence, the isotropic equivalent of the 
emergent FRB luminosity is $B(R_{ns})^2 R_{ns}^3 c/R$, and to produce a FRB of 
luminosity $10^{44}$ erg s$^{-1}$ \alfven waves with amplitude   
$B(R_{ns}) \sim 10^{11}$G are required.

An \alfven wave packet of amplitude 
$B(R_{ns})\sim10^{11}$G becomes charge starved when e$^\pm$ density 
of the medium at radius $R$ falls below $\sim 10^{13}/R_7^3$ cm$^{-3}$.
The plasma frequency in the charge starvation region is estimated to 
be $\sim 30 R_7^{-3/2}$GHz, and clumps that form due to two-stream 
instability have radial width $\sim$ (1 cm) $R_7^{3/2}$. 

Calculations of electric field in the charge starvation region, the
formation and acceleration of $e^\pm$ clumps are presented in \S\ref{alfven}
and \ref{clump-acc}. 
Acceleration of clumps which are subjected to collective radiation reaction 
forces -- because particles in the clump radiate in phase -- is a problem 
that cannot be solved from first principles. 
We have relied on causality and energy conservation arguments 
to model radiation reaction forces and estimate clump Lorentz
factor in the electric field of \alfven waves in charge starvation regions. 
We find that clumps are accelerated by the intense electric field on a time 
scale of $\sim 10$ ns to Lorentz factor $\sim 10^3$.  Particles with this 
LF produce curvature radiation at a frequency of $\sim 1$ GHz.
The observed frequencies and luminosities of FRBs can be understood 
if \alfven waves become charge starved somewhere between a few times 
$R_{ns}$ and $\sim 10^2 R_{ns}$.

The predictions of the model are described in \S\ref{rad-physics}.
We find that the minimum and the maximum frequencies expected for FRBs 
according to this model are $\sim 100$MHz and $\sim10^2$ GHz respectively. 
The minimum FRB luminosity (isotropic equivalent) is $\sim 10^{39} [B_{ns,15} 
\lambda^{aw\perp}_{,4}]^2/(R_7 P_{ns}^2)$ erg s$^{-1}$, which 
is set by the fact that lower luminosity \alfven waves don't become charge 
starved as they travel through the magnetosphere, and hence they don't decay 
and produce coherent EM waves. The beaming angle of FRBs, 
according to this model, is about $10^{-2}$rad, and thus the total energy 
budget of a typical fast radio burst is of order 10$^{36}$ erg.
These predictions are dependent on a few uncertain parameters, the 
primary one of which is the radius where the \alfven wave becomes charge 
starved. These dependencies can be found in appropriate equations 
in \S\ref{rad-physics}.

\section{acknowledgments}

We thank the referee for numerous helpful comments and suggestions, which
clarified many concepts and improved the paper. PK would like to thank 
Bing Zhang and Wenbin Lu for comments and discussions.

\end{document}